# Parameter Estimation of Jelinski-Moranda Model Based on Weighted Nonlinear Least Squares and Heteroscedasticity

Jingwei Liu [1,*], Yi Liu[2], Meizhi Xu [3]

[1] School of Mathematics and System Sciences, Beihang University, Beijing, China, 100191

[2] School of Mathematics and Statistics, Beijing Institute of Technology, Beijing, China, 100081

[3] Department of Mathematics, Tsinghua University, Beijing, China, 100084

**Abstract.** Parameter estimation method of Jelinski-Moranda (JM) model based on weighted nonlinear least squares (WNLS) is proposed. The formulae of resolving the parameter WNLS estimation (WNLSE) are derived, and the empirical weight function and heteroscedasticity problem are discussed. The effects of optimization parameter estimation selection based on maximum likelihood estimation (MLE) method, least squares estimation (LSE) method and weighted nonlinear least squares estimation (WNLSE) method are also investigated. Two strategies of heteroscedasticity decision and weighting methods embedded in JM model prediction process are also investigated. The experimental results on standard software reliability analysis database-Naval Tactical Data System (NTDS) and three datasets used by J.D. Musa demonstrate that WNLSE method can be superior to LSE and MLE under the relative error (RE) criterion.

**Keywords:** failure data, reliability estimators, least squares, maximum likelihood estimation, weighted nonlinear least squares.

## 0. Introduction

Probabilistic modeling and parameter estimation is one of core issue of software reliability in recent four decades [1-8]. Jelinski-Moranda (JM) model [1] is a first probabilistic model or statistical model appeared in the software reliability research field [2-8], which was published by Jelinski and Moranda in 1972.

Although, JM model has many deficiencies on assumptions of its theoretical framework, the revisions and amendments of JM model lead to a series of new software reliability model[2-8], and the discussions have been lasted up to now[9,10]. Literature [10] concluded a series of research studies related to JM model. Meanwhile, the parameter estimate approach in the literature [10] announced that using its modified JM model, the parameter estimation using the forth n=26 failure data of NTDS could get an estimation of the total number of software errors $\hat{N}_0 = 34.2222$, which is relative accurate and consistent with the actual total number of errors. While in the first manuscript of this paper in 2006 with weighted least squares estimation model of JM model, the estimation results of NTDS failure data with n=26 failure data would obtained total number estimation close to $\hat{N}_0 = 34$ under some certain weighting functions (See Table 5).

Literature [6] pointed out that using JM model with the n=26 failure data of NTDS failure data, the parameter estimation of $\hat{N}_0 = 31.2$ is available. However, the real total number of error inherent in the NTDS program is $N_0 = 34$, the estimated value is obviously lower than the actual value. Synchronously, in the actual scientific computing, it is found that the estimated value of the literature [8] for JD Musa three groups of software [2,8] for total error numbers are generally larger than real observed failure number (see 2.2). For NTDS and two sets of JD Musa software reliability data sets,

---

[11] established a chaos software reliability model and elicited that the software reliability cumulative failure time data prediction are effected by chaos or randomness. Its experimental results indicate that NTDS does not have a chaotic phenomenon, but JD Musa two sets of software reliability data have chaotic phenomena. After the chaos preprocessing, the prediction effect with chaos data is better than the data without chaos.

The motivation of this paper is partially from the question proposed by Briand, Basili and Thomas in 1992 about the condition of software reliability data C5 and qualifications necessary conditions weaken R5, which are relative to the statistical software reliability modeling of heteroscedasticity and software reliability prediction performance problem[12]. It is well known that the probability model and statistical model are two different models with different issues to study. Essentially, probabilistic model is based on the probability space of random variable $(\Omega, \mathscr{F}, P)$ [13], whereas the statistical model is based on the statistical structure $(X, \mathscr{B}_X, \mathscr{P}_\Theta)$ of the sample $X$ to estimate the probability family $\mathscr{P}_\Theta$ [14]. For the heteroscedasticity issue raised from C5 and R5 by Briand, Basili and Thomas has not been systematically solved in previous literatures in software reliability field up to now. This paper attempts to addressed it in the JM software reliability modeling framework.

In classical statistics, heteroscedasticity problem is mainly discussed in case of linear regression model, an effective method to overcome heteroscedasticity is the weighted least squares method. This is also the foundation of the topic of this article. The purpose of our investigation on WNLSE based JM model is to start from the statistical analysis to discuss whether heteroscedasticity phenomenon of software reliability data in the JM model framework [12], as well as to answer whether to overcome the heteroscedasticity would improve the predicting performance of JM model.

Parameter estimation of $(N_0, \Phi)$ has been a core issue in JM model. In addition to common MLE and LSE, in LSE framework software reliability, [15] proposed a weighted probability JM model, in which the weight function adopted experiential weight function. [16] discussed the weighted least squares fit with S-N fatigue curve, but it is a linear function of regression function, its weighting function using a given weight function. The weighted least squares method in [17] is the multi-variable least squares linear regression function. [18] used the nonlinear least squares method (NLSE) to discuss the component reliability data fitting problem. This paper attempts to study the effectiveness of the weighted nonlinear least squares estimation in the JM model parameter estimation.

Literature [19] brought forward the principle of software reliability "all conclusions should be supported by the failure data to illustrate the problem". In order to improve the accuracy of the estimated parameters of the model and overcome the heteroscedasticity phenomenon, this paper proposes to estimate JM model parameters with WNLSE, and analyzes the possible conditions of failure numbers inherent in MLE, LSE and WNLSE. On NTDS and three JD Musa software reliability data [2,8,9], the software reliability effectiveness of WNLSE, MLE and LSE are compared. The impacts of empirical weight function, optimal weight function and heteroscedasticity weighting process to JM model prediction accuracy are also discussed. Since LSE of JM model is NLSE, the proposed method in this paper is different from the LSE in the literatures [15-18].

## 1. MLE, LSE and WNLSE of JM model

## 1.1 Brief review of JM model

The basic assumptions for the JM model [2-8] are as follows:

(1) The inherent number of errors $N_0$ in the program is an unknown constant.

(2) All errors in programs are independent, each failure time interval is independent to others. Each detected error is excluded absolutely, each troubleshooting only deals with one error, and no new error is introduced in.

(3) The failure rate is constant in each failure time interval, and proportional to the residual errors. In the i-th test interval, the failure rate function is

$$\lambda(x_i) = \Phi \cdot (N_0 - i + 1), \qquad (1)$$

where $\Phi$ is a proportionality constant. $x_i$ is the i-th time interval starting from the (i-1)-th failure. It obeys the exponential distribution with parameter $\Phi \cdot (N_0 - i + 1)$. The density function of $x_i$ is as follows,

$$f(x_i) = \Phi \cdot (N_0 - i + 1) \exp(-\Phi \cdot (N_0 - i + 1)x_i) \qquad (2)$$

Its reliability function is

$$R(x_i) = \exp(-\Phi \cdot (N_0 - i + 1)x_i) \qquad (3)$$

After the (i-1)-th software failure, software mean time to failure (MTBF) is :

$$MTBF_i = \frac{1}{\Phi(N_0 - i + 1)} \qquad (4)$$

The mean of failures time is:

$$m(t) = N_0[1 - \exp(-\Phi t)] \qquad (5)$$

where $t$ is the cumulative failure time.

Obviously,

$$E(x_i) = \frac{1}{\Phi(N_0 - i + 1)}, \qquad D(x_i) = \frac{1}{\Phi^2(N_0 - i + 1)^2} \qquad (6)$$

## 1.2 Brief review of WLNSE and Heteroscedasticity

Assuming that $(x_1, y_1), ..., (x_n, y_n)$ are the n sample points, the fitting function is $y = f(x, \beta)$, if

$$y_i = f(x_i, \beta) + \varepsilon_i, \quad i = 1, ..., n \qquad (7)$$

where $\beta$ is the parameter vector to be estimated. $\varepsilon_i$ is the residual error item.

Denote $S(\beta) = \sum_{i=1}^{n} [y_i - f(x_i, \beta)]^2$. The parameter $\hat{\beta}$ estimated from

$$\min_{\beta} S(\hat{\beta}) = \min_{\beta} \sum_{i=1}^{n} [y_i - f(x_i, \beta)]^2 \qquad (8)$$

is called non-linear least squares estimation (NLSE).
The parameter $\hat{\beta}$ estimated from

$$\min_{\beta} S_W(\hat{\beta}) = \min_{\beta} \sum_{i=1}^{n} w_i [y_i - f(x_i, \beta)]^2, \qquad (9)$$

is called weighted non-linear least squares estimation (WNLSE). where $w_i > 0, i = 1, ..., n$.
In statistical theory, if the variances of the random variables $x_1, \cdots, x_n$ are different, it is called heteroscedasticity problem.
To solve equation (8) (9), the popularly used algorithms are typically iterative algorithm Newton-Raphson method, Marquardt algorithm, and variable metric method, etc. [20].
In order to facilitate the derivation in classical statistical model, it is always assumed that $\varepsilon_i \sim N(0, \sigma_i^2)$, if $\sigma_i$ is different from each other, it is the heteroscedasticity problem [21] mentioned above. In linear regression model, weighted least squares method (WLS) is one of the main technique to deal with heteroscedasticity, namely finding the appropriate weight function to eliminate the effects of heteroscedasticity [21-24]. Theoretically, the optimal weight of WLS method is $w_i = 1/\sigma_i^2$, although [24] gives the conclusion in case of linear regression, it is easy to prove this conclusion still holds in nonlinear regression case.

There are several methods to test heteroscedasticity, for example, Goldfeld-Guandt test, Breusch-Pagan test, White test, etc. [21, 23]. Goldfeld-Guandt test is involved in the experiments, its test procedure is as follows:

The null hypothesis is $H_0: \sigma_1 = \sigma_2 = \cdots = \sigma_N$, and alternative hypothesis is $H_1: \sigma_i \neq \sigma_j$.

1) Rank the observations of $x$ in accordance with values from low to high, as the value of $x$ is considered to be related to the variance.

2) Divide the observations into 3 subgroups, the middle subgroup of $d$ size proportion to 1/5~1/4 of total sample number is omitted. The subgroup with low values and high values are denoted as subgroup 1 and 2 each of $(N-d)/2$ observations.

3) Fit investigated regression model $Y$ on subgroup 1 and 2 respectively, each regression model has $(N-d)/2$ observations and $[(N-d)/2]-k$ degree freedom.

4) Calculate the residual sum of squares on subgroup 1 and 2, denote as $EER_1$ and $EER_2$. And compute the ratio

$$\lambda = \frac{EER_2/((N-d-2k)/2)}{EER_1/((N-d-2k)/2)} \sim F((N-d-2k)/2,(N-d-2k)/2)) \tag{10}$$

where $k$ is the dimension of the model parameters $\beta$.

For a given level of significance $\alpha$, if the calculated $\lambda$ is greater than the critical value of statistic $F$ at the chosen level of significance $\alpha$, the null hypothesis that there is no heteroscedasticity phenomenon can be rejected. Otherwise, there is no heteroscedasticity phenomenon. The significance level $\alpha$ is set to 0.05.

## 1.3 MLE, LSE and WNLSE

Assuming that the time intervals of failure are $x_1,...,x_n$, the likelihood function of $N_0$ and $\Phi$ in JM model is

$$L(N_0,\Phi) = \prod_{i=1}^{n}\Phi(N_0-i+1)\exp\{-\Phi(N_0-i+1)x_i\} \tag{11}$$

Maximizing the log-likelihood function, we obtain

$$\begin{cases} \dfrac{\partial \ln L(N_0,\Phi)}{\partial N_0} = 0 \\ \dfrac{\partial \ln L(N_0,\Phi)}{\partial \Phi} = 0 \end{cases} \tag{12}$$

The MLE estimation of JM model satisfies the following formula

$$\begin{cases} \Phi = \dfrac{n}{N_0(\sum_{i=1}^{n}x_i) - \sum_{i=1}^{n}(i-1)x_i} \\ \sum_{i=1}^{n}\dfrac{1}{(N_0-i+1)} = \dfrac{n}{N_0 - \dfrac{1}{\sum_{i=1}^{n}x_i}(\sum_{i=1}^{n}(i-1)x_i)} \end{cases} \tag{13}$$

LSE is to minimize the objective function

$$S(N_0,\Phi) = \sum_{i=1}^{n}(x_i - \frac{1}{\Phi(N_0-i+1)})^2 \tag{14}$$

Let $\begin{cases} \partial S/\partial N_0 = 0 \\ \partial S/\partial \Phi = 0 \end{cases}$, we obtain

$$\begin{cases} \Phi = \sum_{i=1}^{n}\dfrac{1}{(N_0-i+1)^2} \Big/ \sum_{i=1}^{n}\dfrac{x_i}{(N_0-i+1)} \\ (\sum_{i=1}^{n}\dfrac{x_i}{(N_0-i+1)^2})(\sum_{i=1}^{n}\dfrac{1}{(N_0-i+1)^2}) = (\sum_{i=1}^{n}\dfrac{x_i}{(N_0-i+1)})(\sum_{i=1}^{n}\dfrac{1}{(N_0-i+1)^3}) \end{cases} \quad (15)$$

According to formula (6), since the variances of failure interval vary with the failure intervals, the failure data, obeying or sampling from the JM model, would have to be heteroscedasticity. The classic LSE formula (14) can be viewed as the nonlinear regression with the regression function $x_i = f(i) = \dfrac{1}{\Phi(N_0-i+1)}$. As residual series $\varepsilon_i = x_i - \dfrac{1}{\Phi(N_0-i+1)}$ are definitely not obeying normal distribution and Goldfeld-Guandt test equation (9) has the assumption of the residuals normal distribution, the classic heteroscedasticity determining method mentioned in section 1.2 can not be directly applied to the theoretical JM model framework. This paper attempts to use Goldfield-Guandt with JM software reliability model based on a reasonable explanation: the actual software reliability dataset is more complicated, but it is really obtained in the scientific practice observation. The purpose of software reliability research work is to try to describe the objective random phenomenon. Goldfield-Guandt test is also an attempt to explore whether it is helpful in predicting the performance of failure data for JM model in the case of residuals with normal distribution. From the above analysis, it can be seen that classic JM model and heteroscedasticity determination technique are essentially two different models. However, if only the optimal value of least squares is considered, there is no constraints with normal distribution. Therefore, the combination of Goldfield-Guandt test with JM model is theoretically feasible.

To investgate the heteroscedasticity phenomenon in failure intervals, the weighted nonlinear least squares estimation (WNLSE) is employed in, the objective function is modified to

$$S_W(N_0,\Phi) = \sum_{i=1}^{n} w_i (x_i - \dfrac{1}{\Phi(N_0-i+1)})^2, \quad (16)$$

where $w_i > 0, i=1,...,n$. Minimizing $S_W(N_0,\Phi)$, namely $\begin{cases} \partial S_W/\partial N_0 = 0 \\ \partial S_W/\partial \Phi = 0 \end{cases}$, we can obtain (see Appendix A)

$$\begin{cases} \Phi = \sum_{i=1}^{n}\dfrac{w_i}{(N_0-i+1)^2} \Big/ \sum_{i=1}^{n}\dfrac{w_i x_i}{(N_0-i+1)} \\ (\sum_{i=1}^{n}\dfrac{w_i x_i}{(N_0-i+1)^2})(\sum_{i=1}^{n}\dfrac{w_i}{(N_0-i+1)^2}) = (\sum_{i=1}^{n}\dfrac{w_i x_i}{(N_0-i+1)})(\sum_{i=1}^{n}\dfrac{w_i}{(N_0-i+1)^3}) \end{cases} \quad (17)$$

The formula form is concise, especially for the case of a given weight, the solution can be calculated as those of MLE and LSE easily, the Newton-Raphson iterative method is adopted for solving.

From the previous discussion and formula (6), apparently, the optimal weight function is

$$w_i = 1/(\dfrac{1}{\Phi^2(N_0-i+1)^2}) = \Phi^2(N_0-i+1)^2 \quad (18)$$

According to formulae (13) (15) (17), Denote

$$f_{MLE}(N_0) = \sum_{i=1}^{n}\dfrac{1}{(N_0-i+1)} - \dfrac{n}{N_0 - \dfrac{1}{\sum_{i=1}^{n}x_i}(\sum_{i=1}^{n}(i-1)x_i)} \quad (19)$$

$$f_{LSE}(N_0) = (\sum_{i=1}^{n}\dfrac{x_i}{(N_0-i+1)^2})(\sum_{i=1}^{n}\dfrac{1}{(N_0-i+1)^2}) - (\sum_{i=1}^{n}\dfrac{x_i}{(N_0-i+1)})(\sum_{i=1}^{n}\dfrac{1}{(N_0-i+1)^3}) \quad (20)$$

$$f_{WLS}(N_0)=(\sum_{i=1}^{n}\frac{w_i x_i}{(N_0-i+1)^2})(\sum_{i=1}^{n}\frac{w_i}{(N_0-i+1)^2})-(\sum_{i=1}^{n}\frac{w_i x_i}{(N_0-i+1)})(\sum_{i=1}^{n}\frac{w_i}{(N_0-i+1)^3}) \quad (21)$$

To obtain the accurate estimations of formulae (13) (15) (17), the Newton iterative algorithm is adopted to solve the solutions of $f_{MLE}(N_0)=0$, $f_{LSE}(N_0)=0$ and $f_{WLS}(N_0)=0$ in formulae (19) (20) (21). The control accuracy is set to $10^{-16}$. Then, the solutions of $N_0$ are substituted into the formulae (13) (15) (17) respectively, to obtain the corresponding $\Phi$ value.

## 1.4 WNLSE weight function selection and heteroscedasticity process

Weighting function selection is the key issue of WNLSE. Suppose $\{x_1,\cdots,x_k\}$ is a segment of failure data ($2<k\leq n$). Three types of weighting methods are taken into account for the selection of the software reliability problem: empirical weight function, optimal weight function and heteroscedasticity process approach.

Empirical weight function is constructed by time and number of failures, eight empirical weight functions are as follows:

$$w_i = (\frac{\sum_{j=1}^{i} x_j}{i}), \quad i=1,...,n. \quad (21)$$

$$w_i = (\frac{i}{\sum_{j=1}^{i} x_j}), \quad i=1,...,n. \quad (22)$$

$$w_i = i^{\beta}, \quad i=1,...,n. \quad \beta>0 \quad (23)$$

$$w_i = i^{-\beta}, \quad i=1,...,n. \quad \beta>0 \quad (24)$$

$$w_i = i, \quad i=1,...,n. \quad (25)$$

$$w_i = 1/i, \quad i=1,...,n. \quad (26)$$

$$w_i = \sum_{j=1}^{i} x_j, \quad i=1,...,n. \quad (27)$$

$$w_i = 1/\sum_{j=1}^{i} x_j, \quad i=1,...,n. \quad (28)$$

Substituting the eight weighting function mentioned above into (20) respectively, the WNLSE methods (Fig 1(a)) are denoted as: WNLS-1, WNLS-2, WNLS-3, WNLS-4, WNLS-5, WNLS-6, WNLS-7, WNLS-8, where in WNLS-2, WNLS-3, we set $\beta=0.5$. The computation procedure is as follows: Directly substitute weight functions into formulae (21)(17) to obtain estimation of $(N_0,\Phi)$.

Optimal weight function method is denoted as WNLSopt (Fig 1(b)), weight function is following the formula (18). A uniform calculation procedure for all the failure data is described as follows: First, calculate $(N_0,\Phi)$ with LSE, the optimal weight is adopted as $w_i=\Phi^2(N_0-i+1)^2$, then substitute the optimal weight into formulae (21), and re-estimate the $(N_0,\Phi)$. Equivalently, this method obtains the coarse estimation with LSE and refines $(N_0,\Phi)$ with WNLSE using formulae (21)(17).

Before introducing heteroscedasticity process approach, we describe the Goldfeld-Guandt test on software reliability failure data for heteroscedasticity phenomenon. We take the following procedure:

First, ultilize the LSE to estimate $(N_0, \Phi)$ parameters, calculate $f(i,\beta) = \dfrac{1}{\Phi(N_0 - i + 1)}$, and obtain the regression error $\varepsilon_i = x_i - \dfrac{1}{\Phi(N_0 - i + 1)}$. Then, substitute the regression error $\varepsilon_i$ into formula (9), and calculate the critical value of $F$ distribution at confidence level $\alpha = 0.05$. Finally, determine whether there is heteroscedasticity phenomenon on the failure data.

Both of the two WNLSE methods are developed in the framework of JM model, the classical Goldfeld-Guandt test is not considered on the failure data to verify heteroscedasticity phenomenon in classical statistical sense, that is, whether the segment of failure data passes Goldfeld-Guandt test. While sequentially estimating parameters on $\{x_1, \cdots, x_k\}$ ($2 < k \leq n$), some segments may pass Goldfeld-Guandt test but some segments may not, that is, heteroscedasticity phenomenon may only appear in part of segments failure data according to Goldfeld-Guandt test. A reasonable approach is that if a segment failure data does not pass Goldfeld-Guandt test and has heteroscedasticity phenomenon, we use WNLSE method to estimate parameters, if it has no heteroscedasticity phenomenon, we use classical LSE to estimate parameters. In fact, this process is equivalently to mix two different statistical models and embed into the same segmental failure data for one-step forecasting. For the above mixed process model, we propose two schemes to deal with the heteroscedasticity segment with WNLSE. All of them are uniformly called heteroscedasticity process approach .

**Scheme 1**: Given a segment of failure data, LSE is used to estimate $(N_0, \Phi)$, then Goldfeld-Guandt test is employed to determine whether there is heteroscedasticity. If heteroscedasticity phenomenon is not detected, the previous LSE $(N_0, \Phi)$ is utilized to calculate step prediction of RE; If Goldfeld-Guandt heteroscedasticity phenomenon holds, calculate $f(i,\beta) = \dfrac{1}{\Phi(N_0 - i + 1)}$ and let the optimal weights be $w_i = \Phi^2(N_0 - i + 1)^2$, substitute optimal weights into the formula (21) (17) to re-estimate $(N_0, \Phi)$, and then calculate the RE. This scheme is recorded as WNLS$_{H1}$. This approach corresponds to the formula (18) to select the optimal weight function (Fig 1 (c)).

**Scheme 2**: Given a segment of failure data, use LSE to estimate $(N_0, \Phi)$, then use Goldfeld-Guandt to determine whether there is heteroscedasticity. If heteroscedasticity phenomenon is not detected, the previous LSE $(N_0, \Phi)$ is utilized to calculate step prediction of RE, this step is the same as Scheme 1; If Goldfeld-Guandt heteroscedasticity phenomenon holds, calculate $f(i,\beta) = \dfrac{1}{\Phi(N_0 - i + 1)}$ and $\varepsilon_i = x_i - \dfrac{1}{\Phi(N_0 - i + 1)}$, let the weight be $w_i = \dfrac{1}{\varepsilon_i^2}$, substitute optimal weights into the formula (21) to re-estimate $(N_0, \Phi)$, and then calculate the RE. This scheme is recorded as WNLS$_{H2}$.

The theoretical basis of Scheme 2 is that each observation $\varepsilon_i^2$ could be treated as estimation of $D(\varepsilon_i)$ due to the random dynamic real-time data [14]. In essence, from the definition of probability model JM model, each failure interval data $x_i$ obeys exponential distribution with parameter $\lambda(x_i) = \Phi \cdot (N_0 - i + 1)$, this is a kind of statistic problem along stochastic orbit. The $w_i = \dfrac{1}{\varepsilon_i^2}$ can be reversely compensated to reduce the impact of large variance terms in the weighted least squares objective function. Since $\varepsilon_i = x_i - \dfrac{1}{\Phi(N_0 - i + 1)}$ is obtained from the previous iteration of LSE, consequently, $w_i = \dfrac{1}{\varepsilon_i^2}$ is used as the weight and substituted into WNLSE iteration, this scheme can also be referred to as dynamic iterative weighting approach (Fig 1 (d)).

Naturally, heteroscedasticity process approaches embed classical Goldfeld-Guandt test in sequential JM model predictions, and expect to obtain desirable performance.

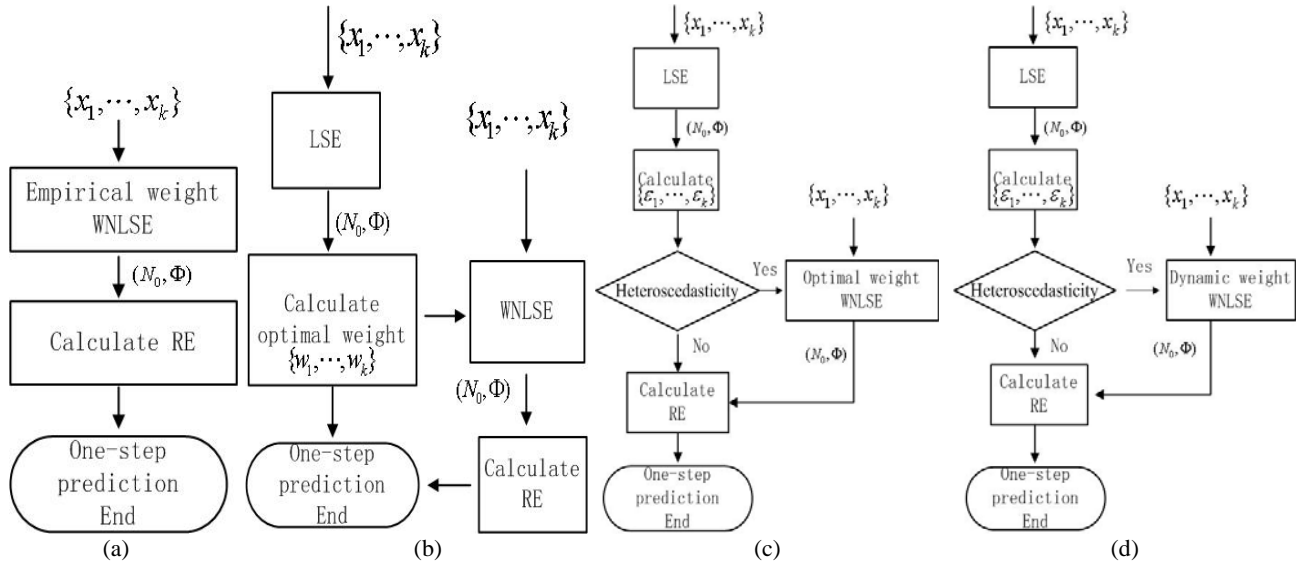

Fig.1 Framework of WNLSE. (a) empirical weight WNLSE (b) optimal weight WNLSE (c) heteroscedasticity optimal weight WNLSE (d) heteroscedasticity dynamic weight WNLSE

## 1.5 Optimal solution of $N_0$

Obviously, the premise of solving the number $N_0$ of errors inherent with MLE, LSE and WNLSE in $f_{MLE}(N_0)=0$, $f_{LS}(N_0)=0$, $f_{WLS}(N_0)=0$ must satisfy $N_0 > n$. To ensure that $N_0$ solutions should exist using Newton-Raphson interative method, the three equations must also be required that their first order, second order derivatives $f'_{MLE}(N_0)$, $f''_{MLE}(N_0)$, $f'_{LSE}(N_0)$, $f''_{LSE}(N_0)$, $f'_{WLS}(N_0)$, $f''_{WLS}(N_0)$, should keep constant sign in their respective neighborhoods of solution $N_0$. Therefore, the optimal solution does not necessarily exist.

Due to the randomness of failure data, even on the same failure data set, the same parameter estimation model, different size of segments would get different $f(N_0)$. $f_{MLE}(N_0)$ on NTDS data and JD Musa three software reliability data is taken for example to analyze the $N_0$ value case.

For $f_{MLE}(N_0)$ (Fig 2-3) on NTDS data, in n=26 case, $N_0$ exits, while in n=31 case, $N_0$ has no solution, only an approximate solution is obtained.

For $f_{MLE}(N_0)$ on J.D Musa data set-I, both in n=12 and 17 case, $N_0$ has no solution, only an approximate solution is obtained (Fig 4-5).

For $f_{MLE}(N_0)$ on J.D Musa data set-II, in n=12 case, $N_0 = 18.7815$, and in n=15 case, $N_0 = 21.2278$ (Fig 6-7). However, the solutions reported in [8] are as follows: in n = 12 case, $N_0 = 41040708$, and in n = 15 case, $N_0 = 61561064$. Apparently, the solutions in [8] are the approximate solution approaching to the asymptote, far larger than the true solution.

For J.D Musa dataset-III, in n=150 case, $N_0 = 152.0197$, and in n=163 case, $N_0 = 179.7473$ (Figure 8-9). The estimation given in [8] is that in n=150 case, $N_0 = 970275456$, and in n=163 case, $N_0 = 970275456$. Clearly, literature [8] also gives an approximate solution.

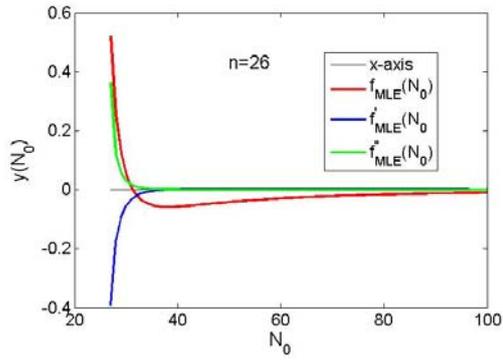

Fig.2 $f_{MLE}(N_0)$ and its first, second derivatives on NTDS with n=26 failure data

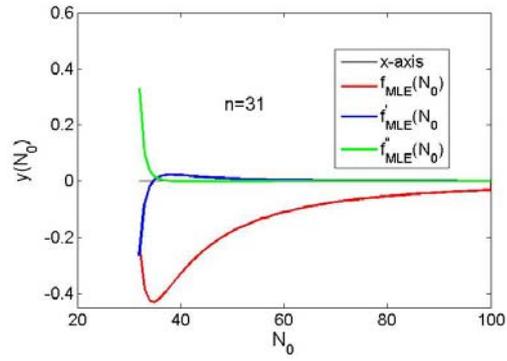

Fig.3 $f_{MLE}(N_0)$ and its first, second derivatives on NTDS with n=31 failure data

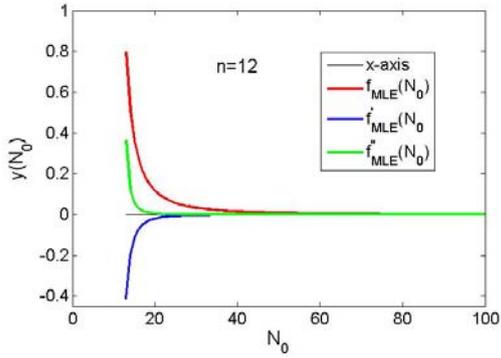

Fig. 4 $f_{MLE}(N_0)$ and its first, second derivatives on J.D Musa-I with n=12 failure data.

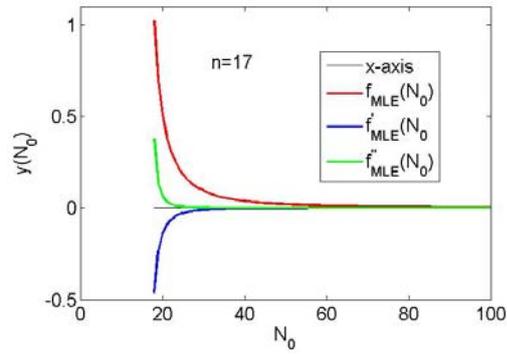

Fig. 5 $f_{MLE}(N_0)$ and its first, second derivatives on J.D Musa-I with n=17 failure data.

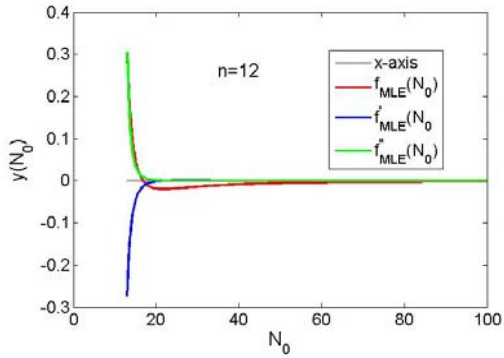

Fig. 6 $f_{MLE}(N_0)$ and its first, second derivatives on J.D Musa-II with n=12 failure data.

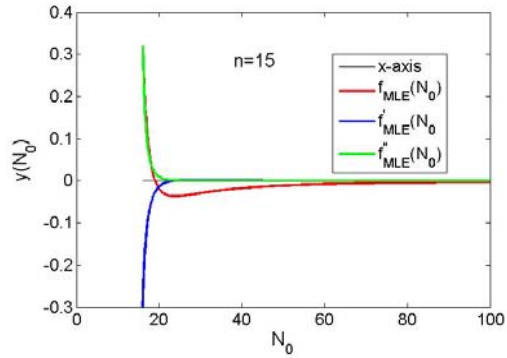

Fig. 7 $f_{MLE}(N_0)$ and its first, second derivatives on J.D Musa-II with n=15 failure data.

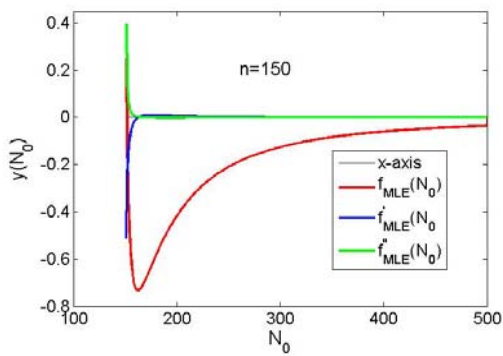

Fig.8 $f_{MLE}(N_0)$ and its first, second derivatives on J.D Musa-III with n=150 failure data.

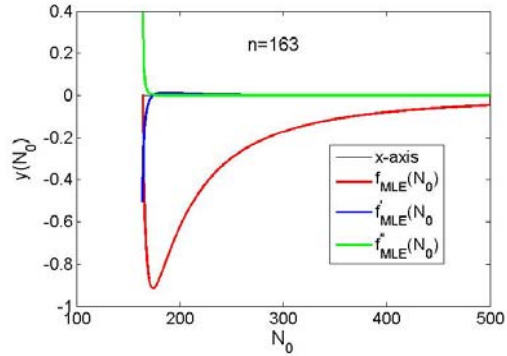

Fig.9 $f_{MLE}(N_0)$ and its first, second derivatives on J.D Musa-III with n=163 failure data.

Similarly, the optimal solutions of $f_{LSE}(N_0) = 0$ and $f_{WLS}(N_0) = 0$ can be analyzed in the same way above, we omit them here. From the several examples above, it can reach conclusion that when estimating the $N_0$ on failure data $\{x_1,\cdots,x_k\}$ （$2 \leq k \leq n$）using MLE,LSE and WNLSE, there are three possible cases:

  1) All optimal solution exist.
  2) All optimal solution does not exist, just obtaining approximate solutions.
  3) Part of segments have optimal solution, and other segments have approximate solution.

Comprehensively, two kinds of optimal solution are proposed for selecting the optimal solution:

  1) According to "all conclusions should be supported by the failure data to illustrate the problem" principle, for any segment of one failure data set, if there is optimal solution, taking the optimal solution for the parameter estimates. If there is no optimal solution, we take the solution obtained by the iteration of Newton's method which meets the accuracy of the optimal solution. For example, Fig 2 and Fig 6-9 satisfy this definition. This method is to take solution derived from the data, hence the solution is called reasonable solution.

  2) The solution satisfying the accuracy of Newton-Raphson algorithm or satisfying the accuracy to the asymptotic is taken as the optimal solution. That is, no matter what the situations shown in Figure 2-9, the approximate solution meeting the accuracy is taken as optimal solution. This solution is called asymptotic solution.

## 2 Failure data sets and model evaluation criteria
## 2.1 Experimental Data

The failure data sets involved in experiments are from literature [2-8], which are widely investigated in [8,9] and international literatures. The failure data sets are composed of classical NTDS data (Table 1) and three sets of software reliability data used by JD Musa (Table 2-4). NTDS is from Jelinski and Moranda (1972), it is the error number and failure intervals of USA Naval Tactical Data System (NTDS) found in the three stages development, testing and client using, there are totally 34 errors, and 26,5,3 in three stages. Here, only the development and testing stages of 31 failure data are involved in experiment.

Table.1  NTDS failure data set (day)

| 9 12 11 4 7 2 5 8 5 7 1 6 1 9 4 1 3 3 6 1 11 33 7 91 2 1 87 47 12 9 135 |
|---|

JD Musa three groups of software reliability data can be found in [2,8], the first set of data consists of 17 failure intervals data, the second set of failure data consists of 15 failure intervals data, the third set is composed of 163 failure interval data. The original data can be found in literature [12] (pp206, pp9, pp456-457). The three datasets is abbreviated as JD Musa-I, JD Musa-II and JD Musa-III.

Table.2 J.D Musa-I failure data set (second)

| 932 3103 661 197 1476 155 1358 288 1169 1061 142 494 660 209 361 688 1046 |
|---|

Table.3  J.D Musa-II failure data set (second)

| 10 9 13 11 15 12 18 15 22 25 19 30 32 25 40 |
|---|

Table.4  J.D Musa-III failure data set (second)

| 320 | 1439 | 9000 | 2880 | 5700 | 21800 | 26800 | 113540 | 112137 | 660 |
|---|---|---|---|---|---|---|---|---|---|
| 2700 | 28793 | 2173 | 7263 | 10865 | 4230 | 8460 | 14805 | 11844 | 5361 |
| 6553 | 6499 | 3124 | 51323 | 17010 | 1890 | 5400 | 62313 | 24826 | 26355 |
| 363 | 13989 | 15058 | 32377 | 41632 | 4160 | 82040 | 13189 | 3426 | 5833 |
| 640 | 640 | 2880 | 110 | 22080 | 60654 | 52163 | 12546 | 784 | 10193 |
| 7841 | 31365 | 24313 | 298890 | 1280 | 22099 | 19150 | 2611 | 39170 | 55794 |

| 42632 | 267600 | 87074 | 149606 | 14400 | 34560 | 39600 | 334395 | 296015 | 177395 |
| 214622 | 156400 | 166800 | 10800 | 267000 | 34513 | 7680 | 37667 | 11100 | 187200 |
| 18000 | 178200 | 144000 | 639200 | 86400 | 288000 | 320 | 57600 | 28800 | 18000 |
| 88640 | 432000 | 4160 | 3200 | 42800 | 43600 | 10560 | 115200 | 86400 | 57600 |
| 28800 | 432000 | 345600 | 115200 | 44494 | 10506 | 177240 | 241487 | 143028 | 273564 |
| 189391 | 172800 | 21600 | 64800 | 302400 | 752188 | 86400 | 100800 | 19440 | 115200 |
| 64800 | 3600 | 230400 | 583200 | 259200 | 183600 | 3600 | 144000 | 14400 | 86400 |
| 110100 | 28800 | 43200 | 57600 | 468000 | 950400 | 400400 | 883800 | 273600 | 432000 |
| 864000 | 202600 | 203400 | 277680 | 105000 | 580080 | 4533960 | 432000 | 1411200 | 172800 |
| 86400 | 1123200 | 1555200 | 777600 | 1296000 | 1872000 | 335600 | 921600 | 1036800 | 1728000 |
| 777600 | 57600 | 17280 | | | | | | | |

## 2.2 Model Evaluation Criteria

Main criteria for software reliabiltiy model evaluation are relative error (RE) of failure interval and the mean time to failure [8], U-plot, Kolmogorov-Smirnov distance [7], etc. To compared with the literature [8,9], the RE criterion is also adopted in our experiments.

$$RE = \frac{1}{n}\sum_{i=1}^{n}\frac{|x_i - MTBF_i|}{x_i} \times 100\% \qquad (30)$$

RE criterion would be divided into two types $RE\_I$ and $RE\_II$, where, $RE\_I$ is defined as follows: for the m failure data m (m <n), construct the parameter model to estimate $(N_0, \Phi)$, then substitute the parameters into formula (4) to calculate $MTBF_i(m)$ of all failure data, finally substitute $MTBF_i(m)$ into formula (30) and calculate RE. In order to verify the capability of established model on the part data would reasonably fit the whole data, and the predictive ability on the rest (n-m) data, This criterion is more reasonable for evaluating a fixed number of errors known case. In order to analyze the predictive capability, the RE criteria on the forth m data and the rest (n-m) data are denotes as $RE\_I\_training$ and $RE\_I\_testing$, which are:

$$RE\_I\_training = \frac{1}{m}\sum_{i=1}^{m}\frac{|x_i - MTBF_i|}{x_i} \times 100\% \qquad (31)$$

$$RE\_I\_testing = \frac{1}{n-m}\sum_{i=m+1}^{n}\frac{|x_i - MTBF_i|}{x_i} \times 100\% \qquad (32)$$

$RE\_II$ criterion is defined as follows: Use $\{x_1, x_2, ..., x_{i-1}\}$ to estimate $N_0, \Phi$, and calculate $MTBF_i$, all $MTBF_i$ are eventually substituted into formula (30) to calculate RE. The purpose of is $RE\_II$ to describe the one-step predictive capability.

The above criteria are the smaller the value, the better the performance. The calculation procedure is same to literature [8]. When n=1,2, the index does not calculate, and all of the above evaluation index start from n=3.

## 3. Results

As in four sets of NTDS, JD Musa-I, JD Musa-II and JD Musa-III, only the failure data NTDS announces in the literature that the fixed failure number is 34. Therefore, three experiments are designed as follows:

**Experiment 1**, The index $RE\_I$ comparison with MLE, LSE and WNLSE on NTDS with m=26 in the case of reasonable solution.

**Experiment 2**, The index $RE\_II$ on all four failure data of NTDS, JD Musa-I, JD Musa-II and JD Musa-III for one-step prediction verify the validity of MLE, LSE and WNLSE in the case of reasonable solution .

**Experiment 3**, The index $RE\_II$ on all four failure data of NTDS, JD Musa-I, JD Musa-II and JD Musa-III for one-step prediction verify the validity of MLE, LSE and WNLSE in the case of asymptotic solution .

The experimental results are shown in Table 5-7.

### 3.1 Criteria comparison of WLE, LSE and WNLSE on NTDS with m=26

For the NTDS data with m=26, when the solution of Newton-Raphson iterative algorithm is set to reasonable solution, the indexes $RE\_I$, $RE\_I\_training$ and $RE\_I\_testing$ of MLE, LSE, WNLSE are as follows:

Table 5. RE_I index of MLE,LSE and WNLSE on NTDS with m=26

|  | N | Φ | RE_I | RE_I Training | RE_I Testing |
|---|---|---|---|---|---|
| MLE | 31.2159 | 0.006849 | 282.4772 | 297.7377 | 203.1224 |
| LSE | 32.0564 | 0.006209 | 282.6287 | 303.9038 | 171.9984 |
| WNLS-1 | **33.2502** | 0.005618 | 278.4294 | 304.3387 | **143.7010** |
| WNLS-2 | 31.0558 | 0.006858 | 288.6679 | 302.7011 | 215.6952 |
| WNLS-3 | 32.7955 | 0.005825 | 279.8486 | 304.3056 | 152.6719 |
| WNLS-4 | 32.3541 | 0.006046 | 281.4133 | 304.1184 | 163.3466 |
| WNLS-5 | **33.0854** | 0.005691 | 278.9254 | 304.3433 | **146.7524** |
| WNLS-6 | 37.7379 | 0.004258 | 268.7858 | **300.9540** | **101.5112** |
| WNLS-7 | **34.9912** | 0.004973 | 274.0887 | 303.5875 | **120.6953** |
| WNLS-8 | 40.1833 | 0.003800 | **265.0097** | 298.3371 | 91.7073 |
| WNLS$_{opt}$ | 31.2159 | 0.006742 | 287.3568 | 302.9925 | 206.0516 |
| WNLS$_{H1}$ | 31.1081 | 0.006819 | 288.1279 | 302.8012 | 211.8266 |
| WNLS$_{H2}$ | 38.5667 | 0.004089 | **267.4298** | **300.0726** | **97.6872** |

Table 5 shows that on NTDS with m=26, WNLS-1, WNLS-5 and WNLS-7 can accurately estimate the intrinsic failure number. In the LSE comparisons , WNLS-6,WNLS-8 and WNLS$_{H2}$ can simultaneously obtain more accurate fitting capibility and predictive performance on training set and testing data than LSE, with the cost of relatively large $N_0$. If only RE index is considered, WNLS-1, WNLS-3, WNLS-4, WNLS-5, WNLS-7 can debase the (n-m) step prediction RE index with the cost of high estimation of $N_0$.

WNLS$_{H2}$ has better performance on both training and testing data sets with RE indes than MLE and LSE. This shows that the integration of classical heteroscedasticity model and JM model has validity on NTDS with m=26 , and its prediction accuracy is better than WNLS$_{H1}$.

The results of WNLS$_{opt}$ suggest that the theoretically optimal WNLSE may not be actually the best in simulation.

The experimental results shows that for NTDS with m=26, the failure number inherent in software reliability could be properly estimated.

### 3.2 Comparison of WLE, LS, WNLS in one step prediction with reasonable solution

Experiment 1 only investigates NTDS with m=26, in this section, NTDS, JD Musa-I, JD Musa-II, JD Musa-III are all involved in one-step prediction with reasonable solution, the $RE\_II$ indexes of MLE, LSE and WNLSE are shown in Table 6.

Table 6. RE_II index of MLE,LSE and WNLSE with reasonable solution on NTDS and three J.D Musa sets

|  | NTDS | J.D. Musa-I | J.D. Musa-II | J.D. Musa-III |
|---|---|---|---|---|
| MLE | 391.5204 | 190.4551 | 20.8767 | 2659.7575 |
| LSE | 314.4524 | 190.5711 | 22.4527 | 1390.0797 |
| WNLS-1 | 744.8887 | **190.4631** | 22.7599 | 1549.4476 |
| WNLS-2 | 344.3282 | 190.9215 | 22.1694 | 2213.2277 |
| WNLS-3 | 378.3873 | 192.6048 | **21.2045** | 1511.5081 |
| WNLS-4 | **309.5991** | 190.5483 | 23.9237 | 2420.5040 |
| WNLS-5 | **301.0724** | 190.5335 | 25.5899 | 1872.7917 |
| WNLS-6 | **275.4452** | 199.1884 | **20.1731** | 1793.8811 |
| WNLS-7 | **275.3428** | **190.4598** | 25.9599 | **1416.6654** |
| WNLS-8 | **289.8090** | 202.5314 | **20.1140** | 6217.7935 |
| WNLS$_{opt}$ | 325.3276 | 190.8644 | **20.8074** | 2175.0450 |
| WNLS$_{H1}$ | **215.4516** | 190.8644 | **20.7737** | 1804.7796 |
| WNLS$_{H2}$ | **254.7178** | 223.5711 | 23.2718 | 1438.4398 |

Table 6 shows that in the view of RE_II criteria, on NTDS both WNLS$_{H1}$ and WNLS$_{H2}$ are superior to MLE and LSE; On JD Musa-I data, the performances of WNLS$_{H1}$ and WNLS$_{H2}$ become worse; On JD Musa-II data, WNLS$_{H1}$ is better than MLE and LSE; On JD Musa-III data, WNLS$_{H1}$ and WNLS$_{H2}$ are superior to MLE and inferior to LSE. On the contrary, on the JD Musa-I and JD Musa-II sets, the empirical weights can be found whose performance is superior to LSE and inferior to MLE.

In Experiment 1, All of MLE, LSE and WNLSE do not change the existence of the optimal solution. In Experiment 2, the existence of optimal solution for one-step prediction is shown in Table 7. Table 7 shows that JD Musa-I and JD Musa-II are two exceptions in the existence of optimal solution, all the optimal solutions of MLE and LSE in one-step prediction on JD Musa-I do not exist, on JD Musa-II all one-step prediction, most of the optimal solutions exist. The above results show that in the case of no optimal solution for software failure data, the weight selection can get slight improvement of prediction accuracy. For NTDS (Table 7), when the solution existence cases occupy the proportions between 6/31 and 11/31, which is about 20%-30%, all of WNLS-4, WNLS-5, WNLS-6, WNLS-7, WNLS-8, WNLS$_{H1}$ and WNLS$_{H2}$ could improve RE_II indexes distinctly. For JD Musa-III, although the proportion of optimal solutions existence is between 110/163 to 153/163, namely 67% to 94%, no suitable strategy could be found to make RE_II index become better. The above experimental results shows that, for the failure data, improvement of RE_II index is not dependent on the number of optimal solutions. Moreover, the results of WNLS$_{H1}$ and WNLS$_{H2}$ in Table 7 suggest that treatment of heteroscedasticity process may not significantly lead to increasing of the number of optimal solutions, for example,on NTDS, JD Musa-I and JD Musa-II. As to JD Musa-III, the changing of optimal solution number depends on the empirical weights.

Table 7. The numbers of optimal solution existed in one step forecast on NTDS and three J.D Musa sets

|  | NTDS | J.D. Musa-I | J.D. Musa-II | J.D. Musa-III |
|---|---|---|---|---|
| MLE | 10 | 0 | 12 | 124 |
| LSE | 7 | 0 | 12 | 122 |
| WNLS-1 | 9 | 0 | 12 | 110 |
| WNLS-2 | 8 | 1 | 12 | 153 |
| WNLS-3 | 7 | 0 | 12 | 121 |
| WNLS-4 | 7 | 0 | 12 | 123 |
| WNLS-5 | 8 | 0 | 12 | 111 |
| WNLS-6 | 7 | 1 | 12 | 151 |
| WNLS-7 | 7 | 1 | 12 | 153 |
| WNLS-8 | 7 | 1 | 12 | 111 |
| $WNLS_{opt}$ | 11 | 0 | 12 | 124 |
| $WNLS_{H1}$ | 6 | 0 | 12 | 122 |
| $WNLS_{H2}$ | 11 | 0 | 12 | 124 |

**3.3 Comparison of WLE, LS, WNLS in one step prediction with asymptotic solution**

In this section of Experiment 3, NTDS,JD Musa-I,JD Musa-II and JD Musa-III are solved by Newton-Raphson algorithm and asymptotic solution strategy is adopted. The $RE\_II$ indexes of one-step prediction of MLE, LSE and WNLSE are as follows:

Table 8. RE_II index of MLE,LSE and WNLSE with asymptotic solution on NTDS and three J.D Musa sets

|  | NTDS | J.D Musa-I | J.D Musa-II | J.D Musa-III |
|---|---|---|---|---|
| MLE | 159.2472 | 190.4539 | 26.6761 | 524.9629 |
| LSE | 159.3476 | 190.5455 | 26.5468 | 525.4689 |
| WNLS-1 | **157.3702** | **190.4617** | 26.6081 | **524.9789** |
| WNLS-2 | 159.7133 | 190.7159 | **26.5347** | 527.0927 |
| WNLS-3 | **157.5956** | 190.5668 | **26.5362** | 526.4024 |
| WNLS-4 | **157.4423** | **190.5311** | 26.5654 | **525.2783** |
| WNLS-5 | **158.4906** | **190.5197** | 26.5830 | **525.1660** |
| WNLS-6 | **157.8555** | 190.5850 | **26.5358** | 526.9570 |
| WNLS-7 | **157.3338** | **190.4590** | 26.6330 | **524.9690** |
| WNLS-8 | **158.1769** | 190.7159 | **26.5347** | 527.0926 |
| $WNLS_{opt}$ | **158.0225** | 190.7159 | **26.5347** | 527.0927 |
| $WNLS_{H1}$ | **158.0219** | 190.7159 | **26.5347** | 527.0927 |
| $WNLS_{H2}$ | **157.3193** | 190.7159 | 26.6135 | 527.0927 |

Table 8 shows that if all solutions of Newton-Raphson algorithms are set to asymptotic solutions, evaluations of various estimation methods basically have no significant fluctuations. FromTable 8, it can be seen that, generally on these four failure data, the proper weights could be found to make RE_II evaluation index obtain better advantages, although this change is subtle sometimes. $WNLS_{H1}$ can make the RE index of NTDS and JD Musa-II become smaller, whereas $WNLS_{H2}$ can make JD Musa-II become smaller. However, these two methods can not improve the fitting performance of JD Musa-I and JD Musa-III.

Compared the results in Table 6 and Table 8, in the case of partial optimal solution existence of NTDS and JD Musa-III, when all the reasonable solution are changed to asymptotic solutions, the model fitting indexes RE_II could be significantly improved. For JD Musa-I, whose reasonable solutions are mostly asymptotic solutions, various empirical weights are ineffective. For JD Musa-II, whose reasonable solutions mostly exist, when optimal solutions change to asymptotic solutions, the one-step predictive ability decreases distinctly.

In Experiment 3, the one-step predictive results of JD Musa-I, JD Musa-II and JD Musa-III with MLE and LSE basically coincide with literature [8], hence it shows that the solutions in [8] are asymptotic solutions. When all the optimal solutions of software failure data exist, such as JD Musa-II, the fitting results of reasonable solutions are better than the asymptotic solutions. Also it can be seen that the solution conditions in [26] consists with the experimental results in the literature [8], all of them discuss the asymptotic solution case.

### 3.4 WNLSE is superior to MLE and LSE

In Experiment 2 with RE criterion, the performance of heteroscedasticity approach $WNLSE_{H1}$ is consistently better than MLE and LSE on NTDS and JD Musa-II.

In Experiment 3 with RE criterion, the performance of heteroscedasticity approach $WNLSE_{H1}$ can be consistently better than the MLE and LSE on NTDS and JD Musa-II.

In order to prove the exsitence of weight function which consistly make WNLSE be superior to MLE and LSE, all the weighting function (22)-(29) are taken as the square form, and denoted as $WNLS^2$-1, ..., $WNLS^2$-8, then the experiments of Experiment 2 and Experiment 3 are repeated with new squared weight functions. Due to paper length limitations, the detail results are omitted.

In Experiment 2, on JD Musa-I and JD Musa-III, when weighting function takes $WNLS^2$-1 form, the RE values of WNLSE on JD Musa-I and JD Musa-III are 190.4539 and 1292.3006 respectively, they are consistently better than RE values of MLE and LSE.

In Experiments 3, on JD Musa-I and JD Musa-III, when weighting function takes $WNLS^2$-7 form, the RE values of WNLSE on JD Musa-I and JD Musa-III are 190.4534 and 524.9623 respectively. Apparently, WNLSE is consistently better than MLE and LSE.

From the above experimental results, in the JM model framework, the appropriate weight function can be find for WNLSE to superior than MLE and LSE. However, through the analysis on experimental simulation results, it can be seen that, the improvement of WNLSE-based JM model would be realized in neither reasonable solution or asymptotic solution, however, in some cases, the improvement is not significant, but subtle. This conclusion is similiar to the conclusion of WLS based linear regression in literature [19]. Moreover, the combination of classic heteroscedasticity test with JM model is only effective in part failure data, not consistent to all. In addtion, the theoretically optimal weight function $WNLS_{opt}$ is also effective in part of the failure data sets.

Because this research can be categorized into the weighted least squares nonlinear regression in statistics. This study not only answers the heteroscedasticity improvement issues of software reliability in the JM framework proposed by Briand, Basili and Thomas in 1992, but also provides software reliability estimation of the parameters and simulations. This article runs on a uniform test platform without deliberate set for different weights for WNLSE. In software reliability analysis, different computing platform may get different calculation, but this difference is slight. In this paer, the aim to list all the experimental results of MLE and LSE is to contrast with domain and international computing platforms, to increase the trustiness of our simulation results.

## 4 Conclusion

This paper proposes a WNLS method to estimate the parameters of JM software reliability model, and analyses the conditions of inherent error number of MLE,LSE and WNLSE, discusses the heteroscedasticity problem of software reliability, and proposes heteroscedasticity process approach and heteroscedasticity embedded JM model for prediction. From the theoretical and practical simulations, we answer the heteroscedasticity problem in software reliability proposed by Briand, Basili and Thomas in 1992. Theoretically, the failure data obeying JM model must have heteroscedasticity phenomenon. It is found that WNLSE can improve the estimatin of inherent error number $N_0$ on NTDS with m=26. And, heteroskedasticity process approach can obtain better performance of WNLSE than MLE and LSE. Furthermore, under RE criterion, the optimal weight function can be found to make WNLSE better than MLE and LSE on all of the 4 failure data sets. Also, it is found that the existence of optimal solution has not definitely impact on the fitting accuracy of the RE model, RE accuracy would depend on the random characteristics of the failure data. Simultaneously, the simulation results of this paper explains the reasonableness of experimental results on some failure data sets by JM model in literature [8]. Further research will focus on heteroscedasticity problem of more software failure reliability data sets, investigate the deep theoretical connotation of the heteroscedasticity problem, discuss the impact mechanisms of randomization, chaos and heteroscedasticity phenomena on model accuracy and propose the solving strategies.

## Acknowledgements

The first manuscript was finished in Feb. 2006. This version is revised during my visiting University of Southern California, USA. The research is partially supported by CHINA SCHOLARSHIP COUNCIL (No. 201303070216) and Major Program of the National Natural Science Foundation of China (No.61327807).

## Appendix A：

$$S_W(N_0,\Phi) = \sum_{i=1}^{n} w_i (x_i - \frac{1}{\Phi(N_0 - i + 1)})^2 ,\text{ where } w_i > 0, i = 1,...,n.$$

Let $\begin{cases} \partial S_W / \partial N_0 = 0 \\ \partial S_W / \partial \Phi = 0 \end{cases}$, we obtain

$$\begin{cases} \dfrac{\partial S_W}{\partial N_0} = \sum_{i=1}^{n} 2w_i (x_i - \dfrac{1}{\Phi(N_0 - i + 1)}) \dfrac{1}{\Phi} \dfrac{1}{(N_0 - i + 1)^2} = 0 \\ \dfrac{\partial S_W}{\partial \Phi} = \sum_{i=1}^{n} 2w_i (x_i - \dfrac{1}{\Phi(N_0 - i + 1)}) \dfrac{1}{(N_0 - i + 1)} \dfrac{1}{\Phi^2} = 0 \end{cases}$$

The above equations are equal to the following one

$$\begin{cases} \sum_{i=1}^{n} \dfrac{w_i x_i}{(N_0 - i + 1)^2} = \sum_{i=1}^{n} \dfrac{w_i}{\Phi(N_0 - i + 1)^3} \\ \sum_{i=1}^{n} \dfrac{w_i x_i}{(N_0 - i + 1)} = \sum_{i=1}^{n} \dfrac{w_i}{\Phi(N_0 - i + 1)^2} \end{cases} \quad (*)$$

According to the second equation in（*），we obtain

$$\Phi = \sum_{i=1}^{n} \frac{w_i}{(N_0 - i + 1)^2} / \sum_{i=1}^{n} \frac{w_i x_i}{(N_0 - i + 1)}$$

Let the two equations in（*）divide each other on each side of equal mark，we obtain

$$(\sum_{i=1}^{n} \frac{w_i x_i}{(N_0 - i + 1)^2})(\sum_{i=1}^{n} \frac{w_i}{(N_0 - i + 1)^2}) = (\sum_{i=1}^{n} \frac{w_i x_i}{(N_0 - i + 1)})(\sum_{i=1}^{n} \frac{w_i}{(N_0 - i + 1)^3})$$